\documentclass[twocolumn,showpacs,amsmath,amssymb,pre,aps]{revtex4}   
\usepackage{graphicx}
\usepackage{dcolumn}
\usepackage{bm}
\usepackage{color}
\usepackage{relsize}

\newif\ifgraph

\graphtrue             %

\begin{document}
\title{
Pseudo-Chemotactic Drifts of Artificial Microswimmers}

\author{Pulak K. Ghosh$^{1}$, Yunyun Li$^{2}$, Fabio
Marchesoni$^{2,3}$ and Franco Nori$^{4,5}$}
 \affiliation{$^{1}$ Department of Chemistry,
Presidency University, Kolkata 700073, India}
 \affiliation{$^{2}$ Center for Phononics and Thermal Energy Science,
 School of Physics Science and Engineering, Tongji University, Shanghai 200092,
 People's Republic of China}
 \affiliation{$^{3}$ Dipartimento di Fisica,
Universit\`{a} di Camerino, I-62032 Camerino, Italy}
\affiliation{$^{4}$ CEMS, RIKEN, Saitama, 351-0198, Japan}
\affiliation{$^{5}$ Physics Department, University of Michigan, Ann
Arbor, Michigan 48109, USA}

\date{\today}

\begin{abstract}
We numerically investigate the motion of active artificial
microswimmers diffusing in a fuel concentration gradient. We observe
that, in the steady state, their probability density accumulates in
the low-concentration  regions, whereas a tagged swimmer drifts with
velocity depending in modulus and orientation on how the
concentration gradient affects the self-propulsion mechanism. Under
most experimentally accessible conditions, the particle drifts toward
the high-concentration regions (pseudo-chemotactic drift). A correct
interpretation of experimental data must account for such an
``anti-Fickian'' behavior.
\end{abstract}
\maketitle

\section {Introduction} \label{intro}

Chemotaxis, defined as the movement of motile cells or organisms in
response to a chemical gradient, is a well-studied phenomenon
\cite{Berg}. Bacteria and other single- or multi-cellular organisms
propel themselves up or down the concentration gradient of a
particular substance in their search for nutrients or to avoid
antagonists. Inspired by chemotaxis in biology, researchers
synthesized artificial microswimmers \cite{Schweitzer,Chen} that can
move in response to a chemical stimulus \cite{SenPRL,SenJACS}. They
showed that Janus particles (JP), in the form of two-faced Au-Pt
colloidal rods that catalyze hydrogen peroxide redox, are attracted
by a hydrogen peroxide source. Under such conditions, JP's act as
molecular ``robots'' and can thus be employed in practical
applications, such as the design of new intelligent drugs
\cite{Sen_rev}. More sophisticated chemical robots have been proposed
that utilize artificial chemotaxis to navigate autonomously
\cite{Lagzi}.

The simplest and, possibly, best established model of self-propulsion
is encoded by the Langevin equations
\cite{Gibbs,Lowen,Bechinger,EPJST}
\begin{eqnarray}\label{LE1}
\dot x &=& v_0\cos \phi +\sqrt{D_0}~\xi_x(t),~~~ \dot y = v_0\sin
\phi +\sqrt{D_0}~\xi_y(t),\nonumber \\ \dot
\phi&=&\sqrt{D_\phi}~\xi_\phi(t),
\end{eqnarray}
where ${\bf r}=(x,y)$ are the coordinates of the swimmer in the
plane, $v_0$ its self-propulsion speed, and $D_\phi$ an orientational
diffusion constant, whose reciprocal, $\tau_\phi$, quantifies the
time-persistency of the particle's Brownian motion. Here, $\xi_i(t)$,
with $i=x,y,\phi$, are zero-mean and delta-correlated Gaussian noises
with $\langle \xi_i(t)\xi_j(0)\rangle=2\delta_{ij}\delta(t)$. For
long observation times $t$, with $t\gg \tau_\phi$, or lengths $l$,
with $l\gg l_\phi\equiv v_0\tau_\phi$, the effective diffusion
constant, $D$, defined by the asymptotic law $\lim_{t\to
\infty}\langle {\bf r}^2(t)\rangle=4Dt$ \cite{Risken}, is
$D=D_0+D_s$, where $D_0$ is due to thermal fluctuations in the
suspension, and $D_s=v_0^2/2D_\phi$ is a (typically) much larger
self-propulsion term, which depends on the chemical composition of
the suspension itself.
\begin{figure}
\centering
\includegraphics[width=7.8cm,height=6.1cm]{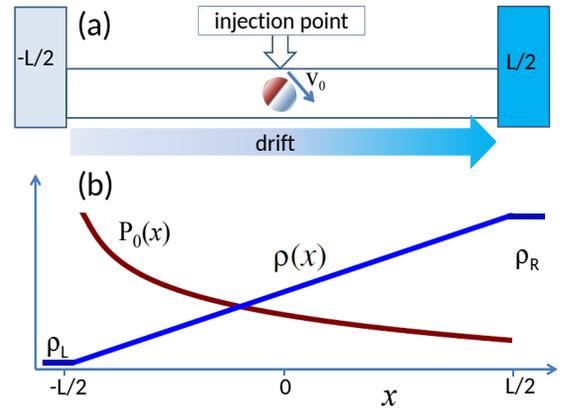}
\caption {(Color online)  Chemical reactor with a stationary fuel
concentration gradient, $\rho(x)$ (see text). A Janus particle
injected in the middle, (a), tends to drift to the right even if its
probability density, $P_0(x)$, peaks on the left (b). The data in (b)
are for $v_0(x) \propto \rho(x)$ with $D_0 = 0.01,\;\eta_v =1,
\delta_v=1, \delta_{\phi}=0 $ and $ D_{\phi} = 0.1$ \label{F1}}
\end{figure}

\begin{figure}[tp]
\centering
\includegraphics[width=7.5cm]{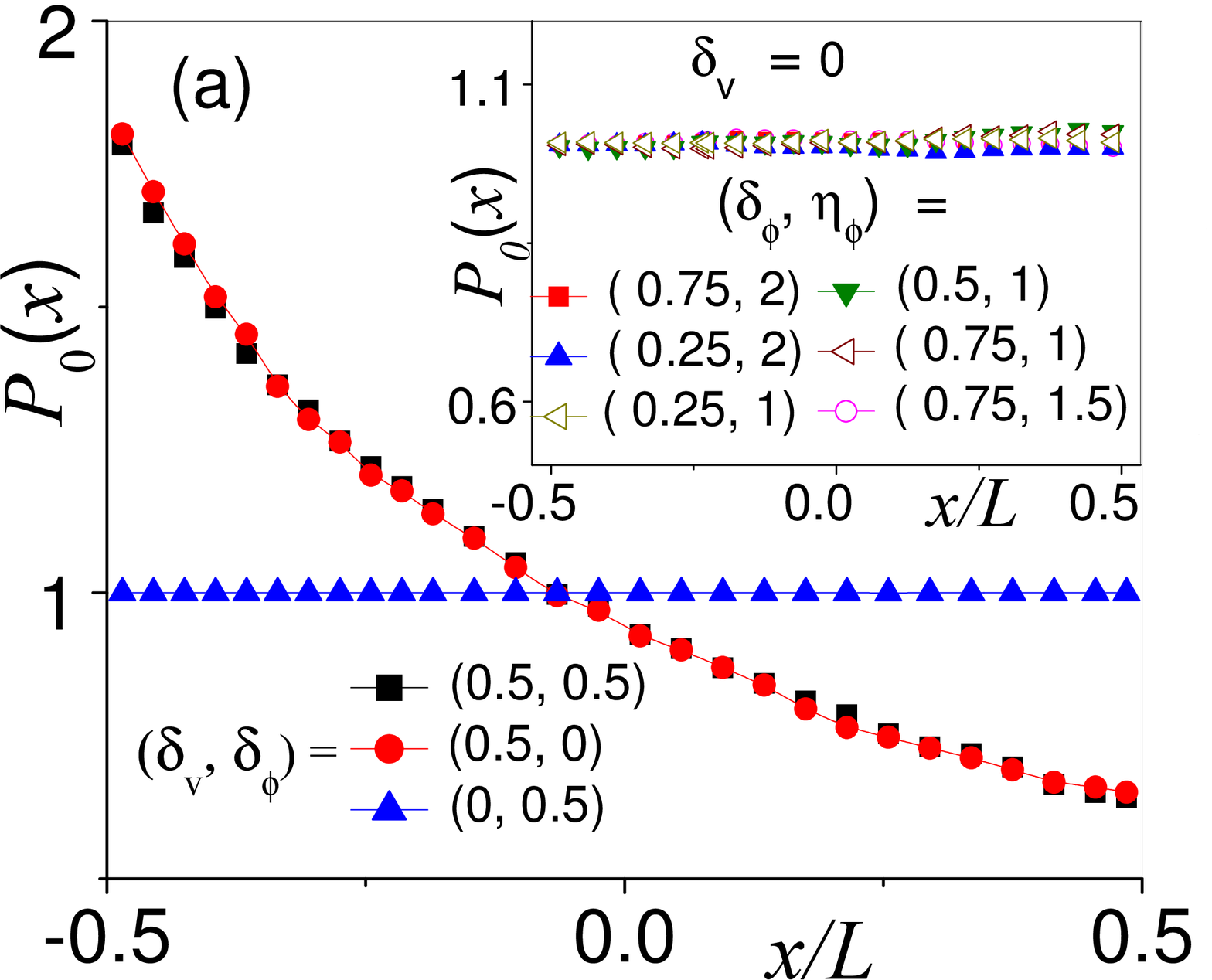}
\vglue 0.5truecm
\includegraphics[width=7.6cm]{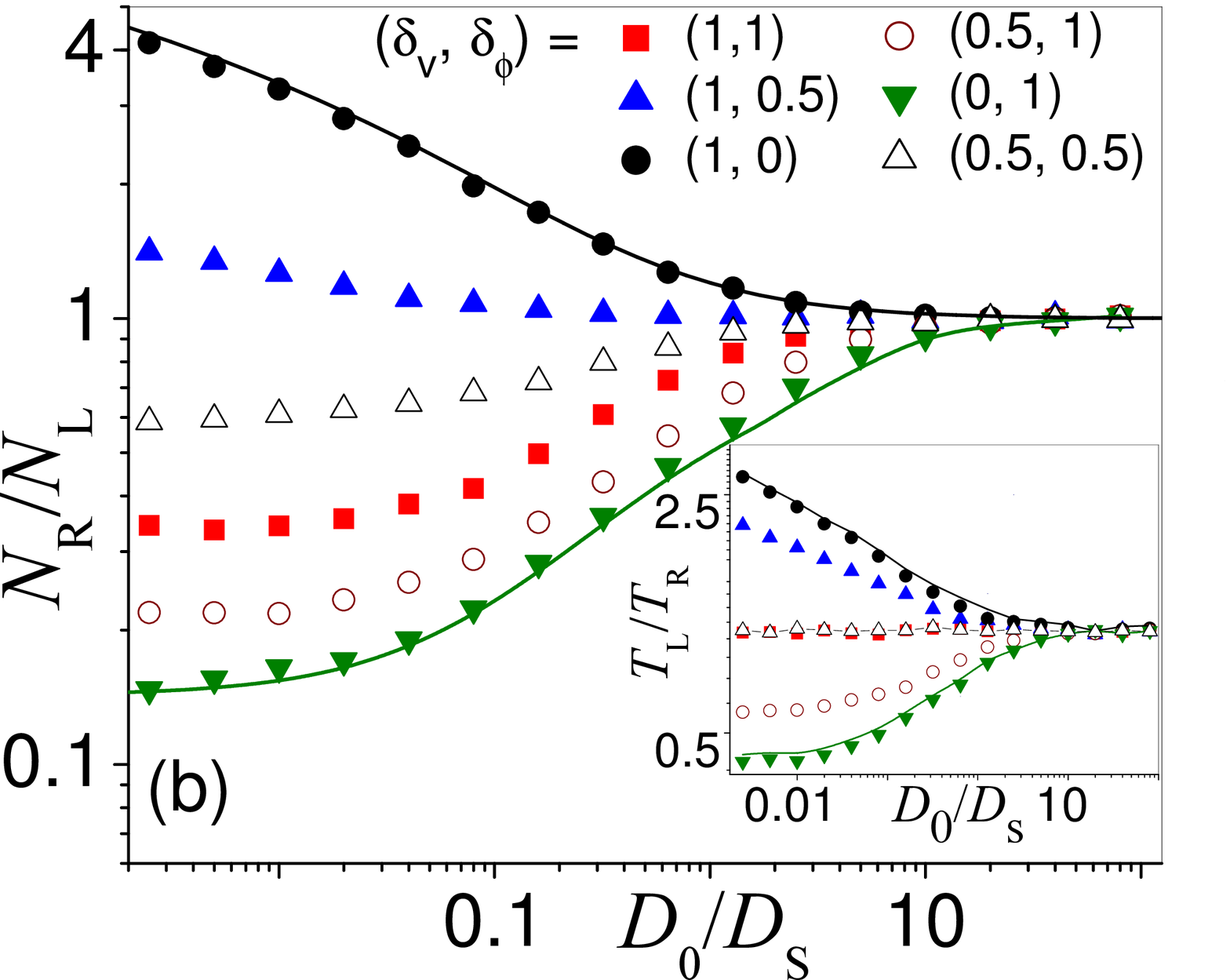}
\vglue 0.5truecm
\includegraphics[width=7.6cm]{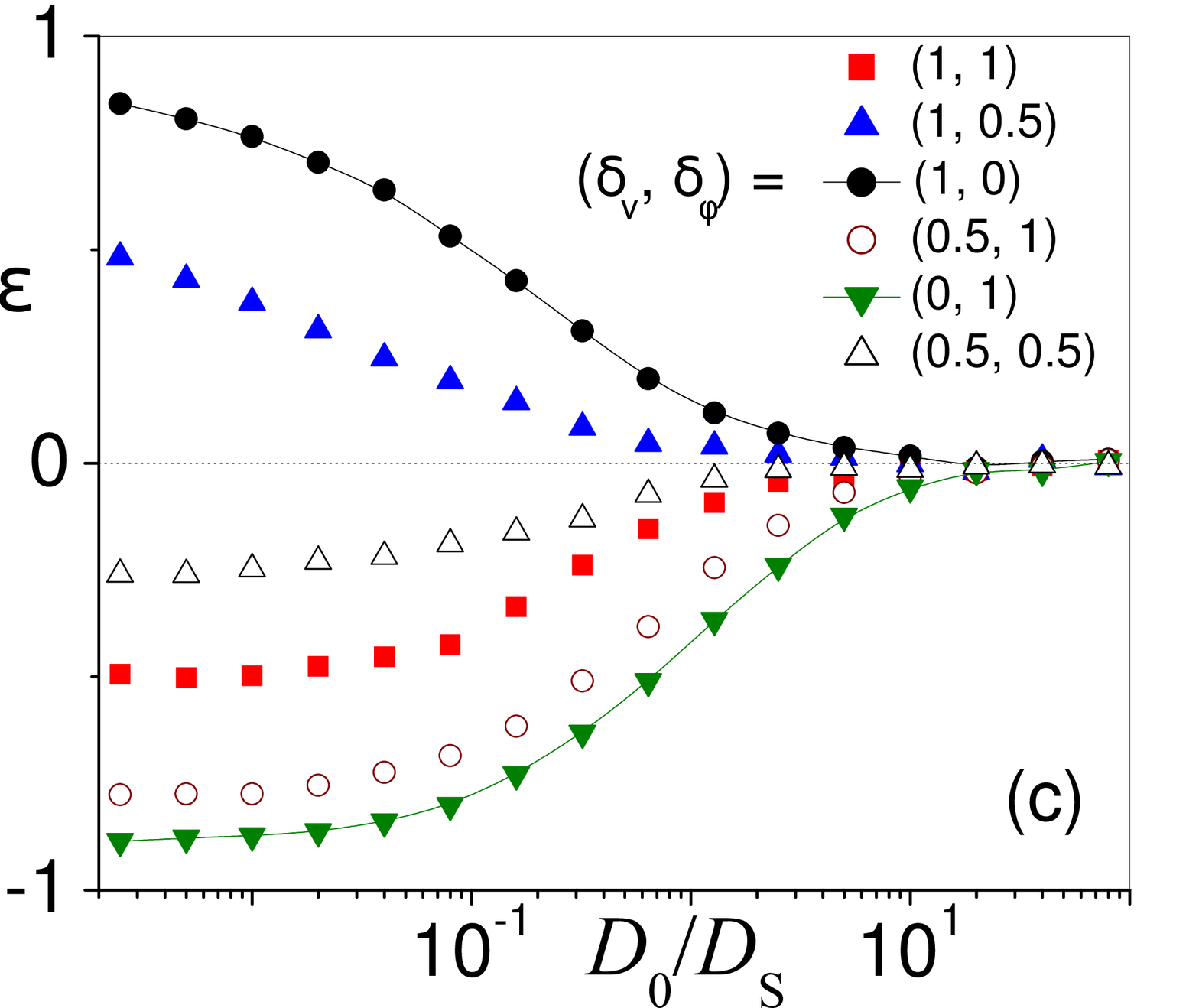}
\caption{(Color online) Janus particle in a concentration gradient,
Eqs. (\ref{x-dep}), with $\eta_v=1$, $\eta_\phi=2$ and different
$\delta_\phi$ and $\delta_v$: (a) $P_0(x)$ for $D_0=0.01$. Data for
$D_0=0.01$ and different $\eta_\phi$ are plotted in the inset; (b)
$N_R/N_L$ and $T_L/T_R$ (inset) vs $D_0$; and (c) $\epsilon$ vs
$D_0$, Eq. (\ref{eps}). Other simulation parameters: $v_0=1$,
$D_\phi=1$, $L=100 l_\phi$ and channel width $y_l=5$. The solid
curves are the analytical predictions based on Eqs.
(\ref{Palpha})-(\ref{NRNL}) with $\alpha=1/2$ and $\alpha=1$,
respectively, for  $\delta_\phi=0$ and $\delta_v=0$. \label{F2}}
\end{figure}

Let us consider now a chemical reactor consisting of a narrow,
straight channel of length $L$ oriented along the $x$-axis, and a
free JP moving in it (Fig.~\ref{F1}) \cite{Fick}. A constant
concentration gradient of the chemical that fuels the particle's
self-propulsion is maintained by connecting the channel to two
reservoirs in thermal equilibrium with concentrations
$\rho_L<\rho_R$. The chemical concentration in the channel,
$\rho(x)$, will then grow linearly from left to right. At the channel
ends, $x=\pm L/2$, two porous membranes allow the chemical flow in
and out, but prevent the JP from escaping into the reservoirs. We
speculate, based on experimental observations
\cite{Gibbs,Ibele,Sen_propulsion,Lugli}, that both $v_0$ and $D_\phi$
(and therefore $D_s$) may depend on $\rho(x)$ to some unspecified
extent. We only assume that both $v_0(x)$ and $D_\phi(x)$ are
non-decreasing functions of the channel coordinate $x$. Indeed, a
higher fuel concentration typically enhances active Brownian motion.
For this reason, the right and left channel endpoints are termed,
respectively, hot and cold. We then ask ourselves two closely related
questions. Upon injecting the JP at the center of the channel, $x=0$:
(1) Which containment membrane is the JP more likely to hit first?
(2) On which side of the channel is it going to sojourn the most
time?

This might sound paradoxical, but we came to the conclusion that the
injected JP is finally attracted toward the left (cold) exit, even
if, immediately after injection, it may drift to either direction,
depending on the $x$-dependence of the propulsion parameters $v_0$
and $\tau_\phi$. For the most common case when the $x$-dependence of
$\tau_\phi$ is much weaker than $D_s$
\cite{Gibbs,Ibele,Sen_propulsion,Lugli}, the injected particle points
decidedly to the right (hot) exit. Reconciling these seemingly
conflicting mechanisms is of paramount importance to control the
chemotaxis of artificial microswimmers as opposed to bacterial
chemotaxis \cite{Schnitzer,Clark}.
To avoid misunderstandings we remark that the adjectives hot and cold
refer here to the regions in the reactor where the effective swimmer
diffusion due to the selfpropulsion, $D_s(x)$, is the highest and
lowest, respectively. The thermal diffusion, $D_0$, is assumed to be
$x$-independent, which means that thermal gradients do not enter our
analysis. Accordingly, in the absence of thermal gradients and for
low fuel concentrations, additional transport contributions due to
hydrodynamic effects in the suspension fluid can be safely neglected.

This paper is organized as follows. In Sec. \ref{numresults} we
present numerical results for the ``splitting probabilities'' that a
JP, injected at the center of the channel, exits it through the right
or left end and the corresponding mean first-exit times. The particle
clearly undergoes a transient drift toward the hot end of the
channel, whereas its stationary distribution tends to accumulate at
the opposite end. In Sec. \ref{discussion} we interpret our data by
means of a phenomenological 1D Langevin equation that describes the
diffusion of a Brownian particle with the spatial dependent diffusion
coefficient $D_s(x)$. The spatial dependence of $D_s(x)$ generates
the drift term here detected as a transient drift. Finally in Sec.
\ref{conclusions} we discuss the implications of our findings in the
interpretation of recent experiments on the diffusion of JP's in
concentration gradients.

\section{Numerical results} \label{numresults}

Our answers to questions (1) and (2) are based on the simulation data
reported in Fig. \ref{F2}. As a study case, we considered the
$x$-dependent self-propulsion parameters,
\begin{eqnarray}\label{x-dep}
v_0(x)=v_0(1+\delta_v x/L)^{\eta_v},~ D_\phi(x)=D_\phi(1+\delta_\phi
x/L)^{\eta_\phi},
\end{eqnarray}
where $\delta_v=\Delta v_0/v_0$ and $\delta_\phi=\Delta
D_\phi/D_\phi$ are both non-negative, and from now on, $v_0$ and
$D_\phi$ are shorthands for $v_0(0)$ and $D_\phi(0)$ at the injection
point. We also set $\eta_v=1$ and $\eta_\phi=2$, so that for
$\delta_v=\delta_\phi$ the self-propulsion diffusion term,
 $D_s(x)=v_0^2(x)/2D_\phi(x)$, is $x$-independent,
i.e., $D_s(x)=D_s$.  That is, the JP is expected to diffuse uniformly
along the channel. On the contrary, we observed that the stationary
probability density function (p.d.f.), $P_0(x)$, of a single JP in
such a close-ended channel tends to accumulate against the left exit,
as displayed in Fig. \ref{F2}(a). This effect is the strongest as
$\delta_v$ is increased at $\delta_\phi=0$. Vice versa, as
$\delta_\phi$ is raised and $\delta_v$ lowered, $P_0(x)$ tends to
flatten out. For $\delta_v=0$, no matter what $\delta_\phi$,
$P_0(x)=L^{-1}$.
Actually, the $x$-dependence of $D_\phi(x)$ seems not to sensibly
affect $P_0(x)$ for any choice of $\eta_\phi$ [Fig. \ref{F2}(a),
inset]. In conclusion, to answer question (2), the injected JP tends
to dwell where $v_0(x)$ is the lowest, as suggested in Ref.
\cite{Schnitzer}, that is by the cold extremity of the channel
(reverse chemotaxis).

\begin{figure}[tp]
\centering
\includegraphics[width=7.5cm]{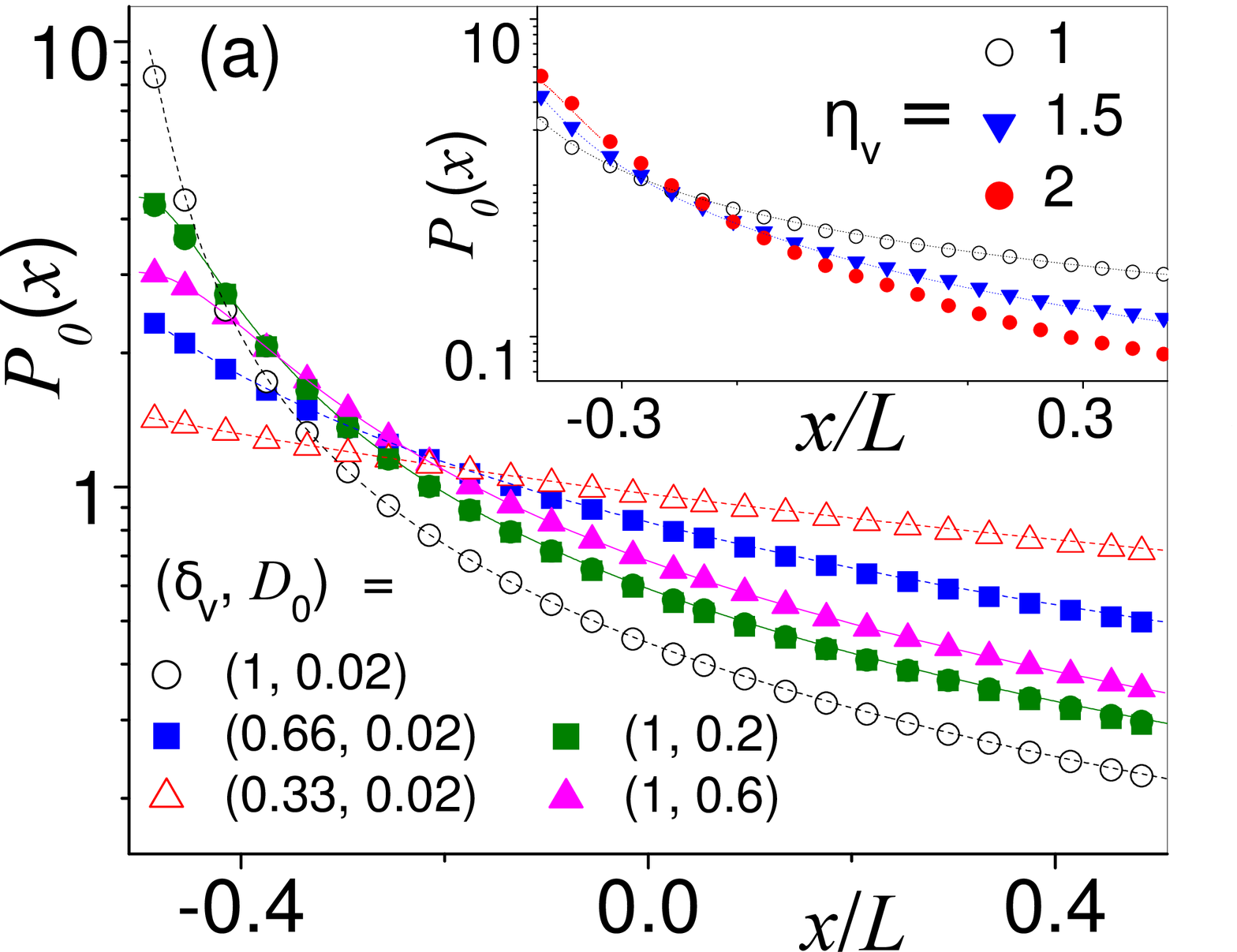}
\vglue 0.5truecm
\includegraphics[width=7.5cm]{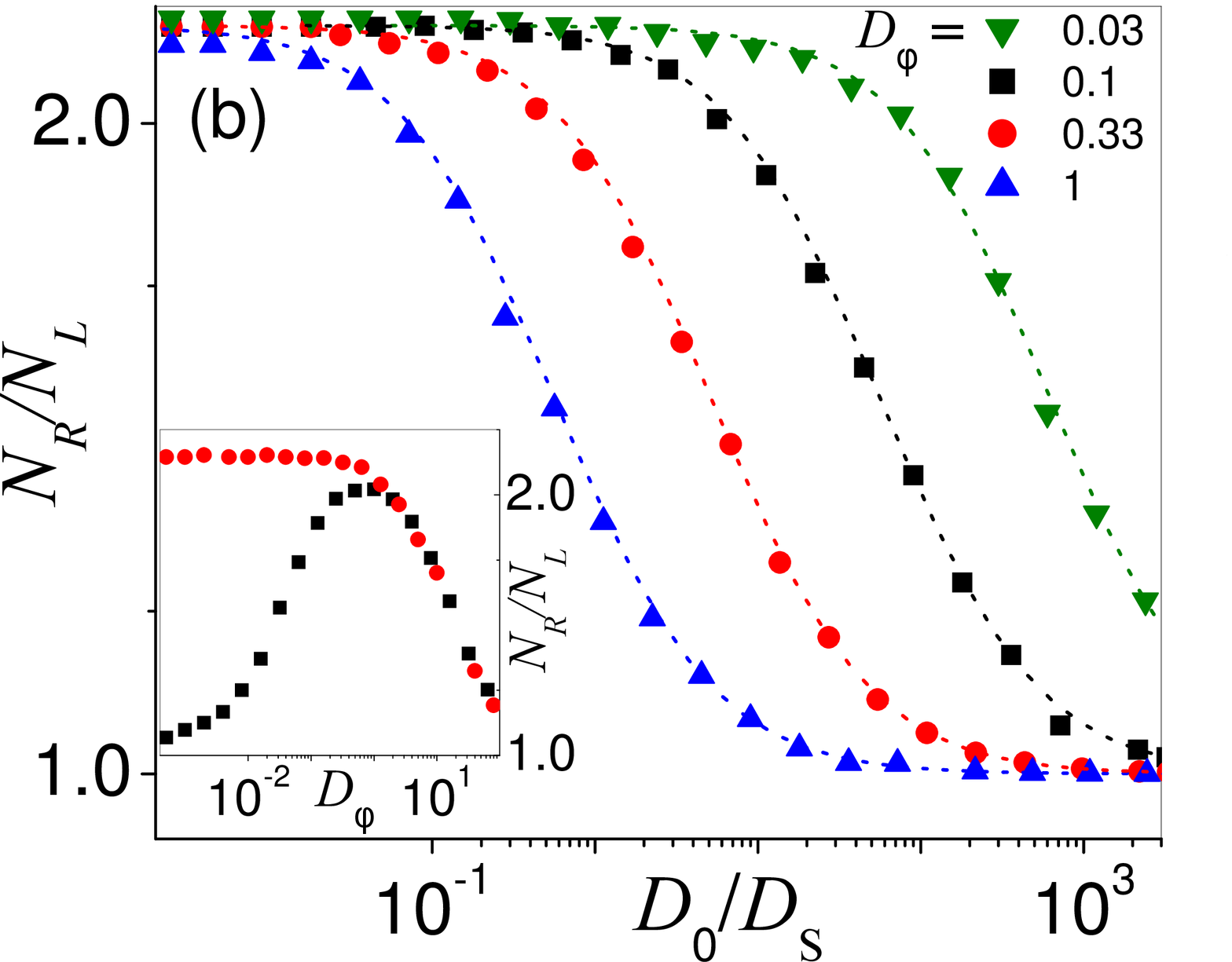}
\vglue 0.5truecm
\includegraphics[width=7.5cm]{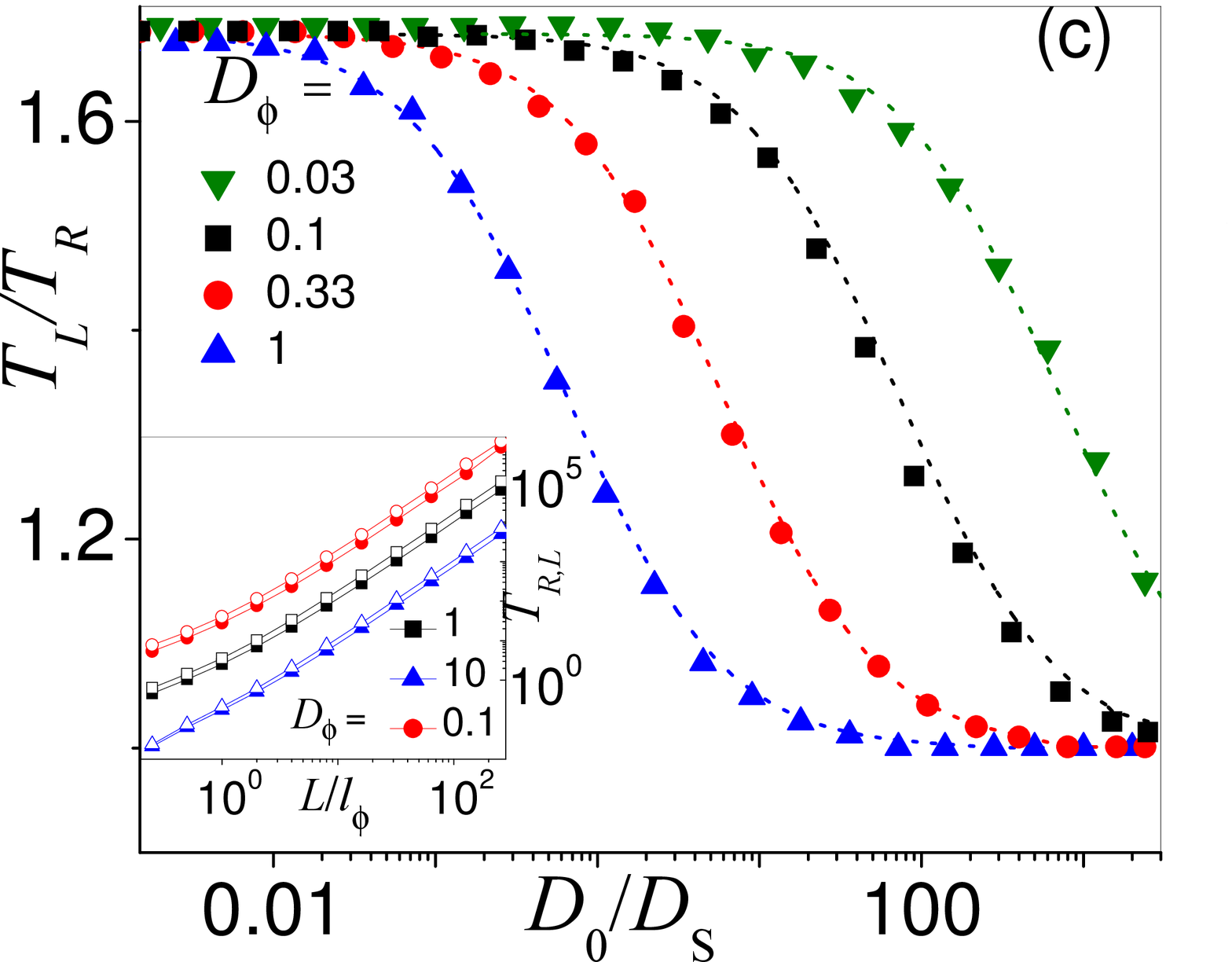}
\caption{(Color online) Channel diffusion for $\eta_v=1$ and
$\delta_\phi=0$: (a) $P_0(x)$ for $D_\phi=0.1, v_0 = 1.5$,  and
different $\delta_v$ and $D_0$ (see legends). Inset: $P_0(x)$ for
$D_0=0.02$, $D_\phi=0.1, \delta_v = 1$ and different $\eta_v$; (b)
$N_R/N_L$ vs. $D_0$ for $\delta_v=2/3$ and different $D_\phi$; and
(c) $T_L/T_R$ vs $D_0$ for $\delta_v=2/3$ and different $D_\phi$.
Other simulation parameters are: $v_0=1.5$, $L=100 l_\phi$ and
channel width $y_l=5$. Inset in panel (b): $N_R/N_L$ vs $D_\phi$ for
$D_0=0.03$ and $L=100$ (squares) and $100 l_\phi$ (circles). Inset in
panel (c): $T_L/T_R$ vs $L/l_\phi$ for $D_0=0.03$, and different
$D_\phi$. The remaining simulation parameters are as in the relevant
main panel. Dashed and solid curves are the corresponding analytical
predictions based on Eqs. (\ref{Palpha})-(\ref{NRNL}) for
$\alpha=1/2$. \label{F3}}
\end{figure}

When one looks at the transient dynamics immediately following the
particle injection, a surprising outcome appears. We injected the
particle at $x=0$ and clocked the time it takes to hit either the
right or left containment membrane. We repeated this numerical
experiment $N=10^6$ times and determined the probability the particle
first reached the right or left exit, $N_{R,L}/N$, and the
corresponding mean-first-passage times (MFPT), $T_{R,L}$, from $0$ to
$\pm L/2$. The ratios $N_R/N_L$ and $T_L/T_R$ are plotted,
respectively, in the main panel and the inset of Fig. \ref{F2}(b). In
the regime of low thermal noise, $D_0\ll D_s$, we obtained distinct
results, depending on which $x$-dependence is stronger, $v_0(x)$ or
$D_\phi(x)$. [Note that we used the same $\eta_v$ and $\eta_\phi$ as
in the main panel (a) for $P_0(x)$.] In the first case, the particle
tries to leave the channel through the right exit and, accordingly,
$T_L>T_R$. Vice versa, on suppressing the $x$-dependence of $v_0(x)$,
while leaving $D_\phi(x)$ unchanged, the particle directs itself
preferably toward to left exit and $T_L<T_R$.

This means that for $\delta_v \gg \delta_\phi$ the injected particle
initially drifts up the $\rho(x)$ gradient (pseudo-chemotaxis), at
odds with Ref.~\cite{Schnitzer}. Only when the increase of $v_0(x)$
along the channel is accompanied by a suitably stronger increase of
its orientational rate, $D_\phi$, the injected particle drifts
immediately down the gradient, in agreement with Fick's law for
ordinary Brownian motion. Magnitude and orientation of the transient
drift are characterized in the forthcoming section by means of the
unique rectification factor $\epsilon$. This result is remarkable
because $P_0(x)$ tends to accumulate in any case around the
concentration minima. This behavior is reminiscent of the ``drift
without current'' effect experimentally observed in Ref.
\cite{Ostrowsky,exp} and numerically investigated in
Refs.~\cite{Tupper} for thermal Brownian motion in confined
geometries. However, the magnitude of the phenomenon reported here is
significantly larger and more easily accessible to experimental
demonstration.

\section{Phenomenological analysis} \label{discussion}

An analytical treatment of the model of Eqs. (\ref{LE1}) is viable in
two limiting cases, i.e., $\delta_\phi=0$, $\delta_v>0$, and
$\delta_v=0$, $\delta_\phi>0$. For this purpose we implemented the
approach of Ref.~\cite{Materials} to reduce the fully
three-dimensional dynamics of Eq. (\ref{LE1}) to the more tractable
1D phenomenological diffusion law,
\begin{equation} \label{LE2}
\dot x= \alpha D'_\alpha(x) +\sqrt{D_\alpha(x)}~\xi(t),
\end{equation}
where the prime sign denotes an $x$ derivation and (i) $\alpha=1/2$
and $D_{1/2}(x)=D_0+v_0^2(x)/2D_\phi$, for $\delta_\phi=0$, and (ii)
$\alpha=1$ and $D_{1}(x)=D_0+v_0^2/2D_\phi(x)$, for $\delta_v=0$.
Here, the multiplicative noise term has to be handled according to
Ito's prescription \cite{Risken} and $\xi(t)$ is defined like the
noises of Eq.~(\ref{LE1}). Note that the Eq.~(\ref{LE2}) can be
rewritten as $\dot x=\sqrt{D_\alpha(x)} \circ \xi(t)$, with $\circ$
denoting the Stratonovitch or anti-Ito prescription, respectively, in
case (i) and (ii). The corresponding Fokker-Planck equation (FPE) is
\begin{eqnarray} \label{FPE}
\frac{\partial}{\partial t}P(x,t)&=&\frac{\partial}{\partial x}\left[
-v_\alpha (x)+\frac{\partial}{\partial x} D_\alpha(x) \right
]P(x,t)\\ \nonumber &=&-\frac{\partial}{\partial x}j(x,t),
\end{eqnarray}
with $v_\alpha(x)=\alpha D'(x)$ for the appropriate value of $\alpha$
\cite{Schnitzer,Marchesoni}. The stationary p.d.f. for zero net
current, $j_0\equiv \lim_{t\to \infty}j(x,t)=0$, reads
\begin{equation} \label{Palpha}
P_0(x)=\lim_{t\to \infty}P(x,t)={\cal N}/D_\alpha(x)^{1-\alpha},
\end{equation}
where ${\cal N}$ is a normalization constant. In particular, for
$\alpha=1$, i.e., $x$-independent $v_0$, $P_0(x) = 1/L$. The
extension to cases with $j_0\neq 0$ is straightforward.

\begin{figure}[tp]
\centering
\includegraphics[width=7.0cm]{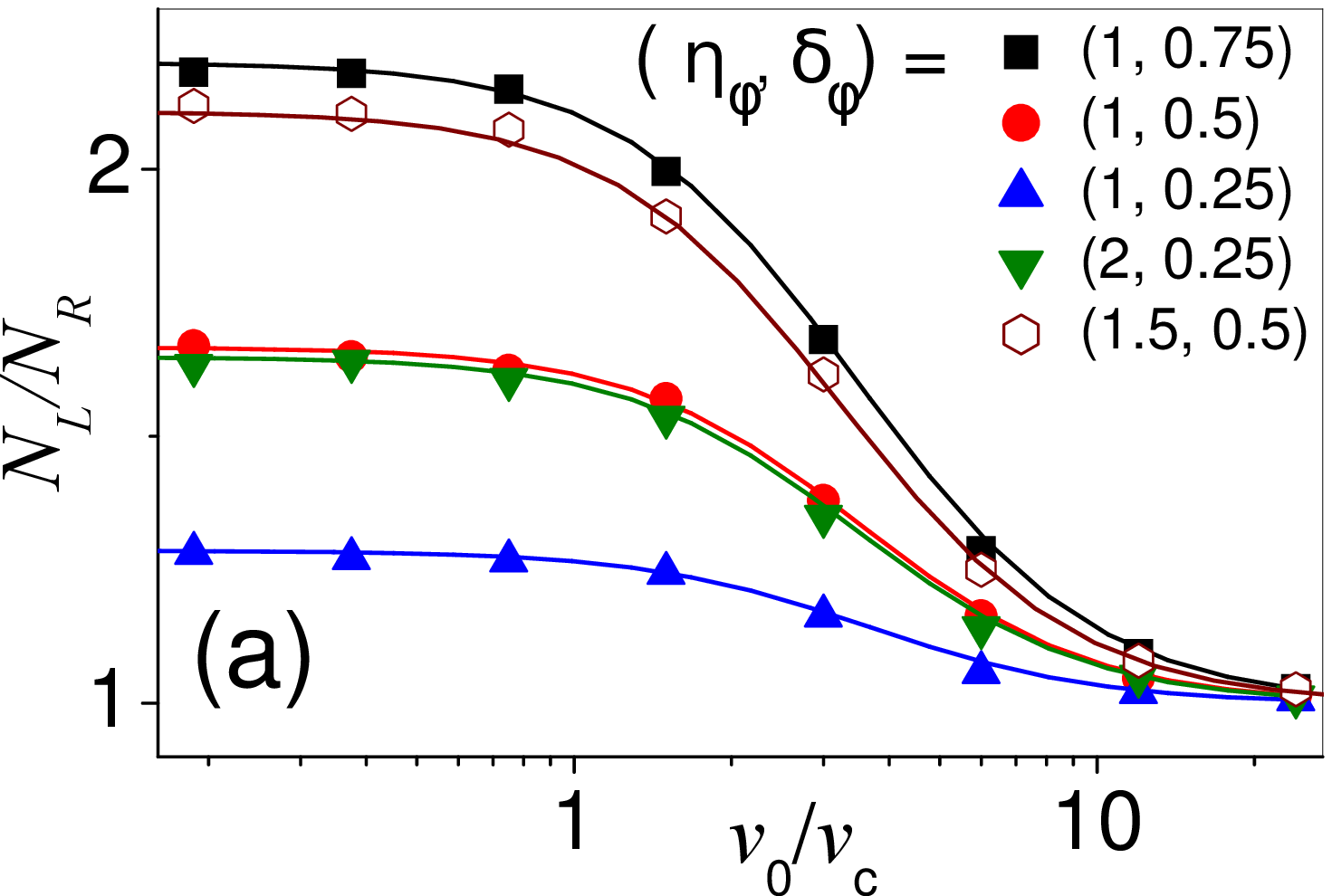}
\vglue 0.5truecm \centering
\includegraphics[width=7.4cm]{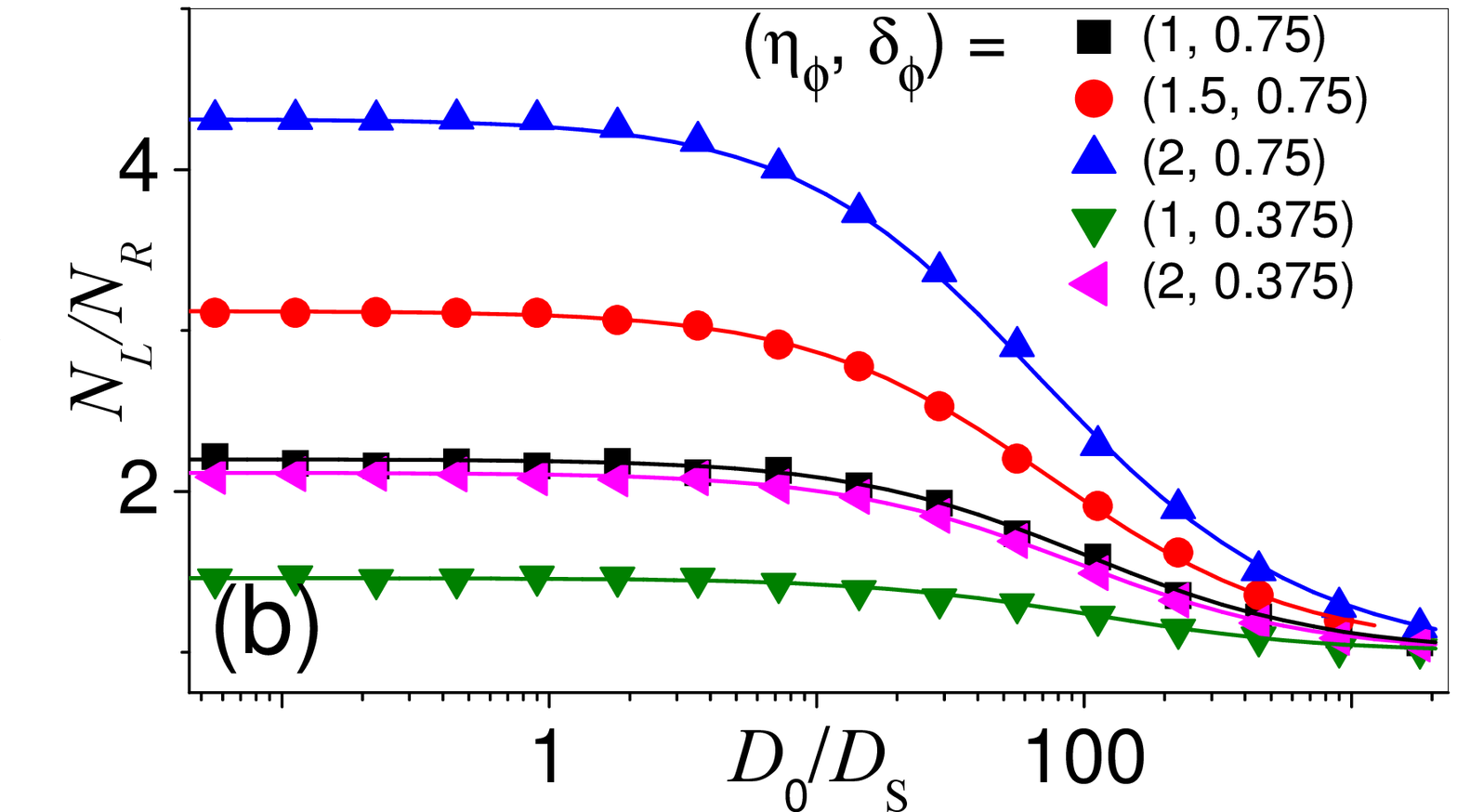}
\vglue 0.5truecm \centering
\includegraphics[width=7.6cm]{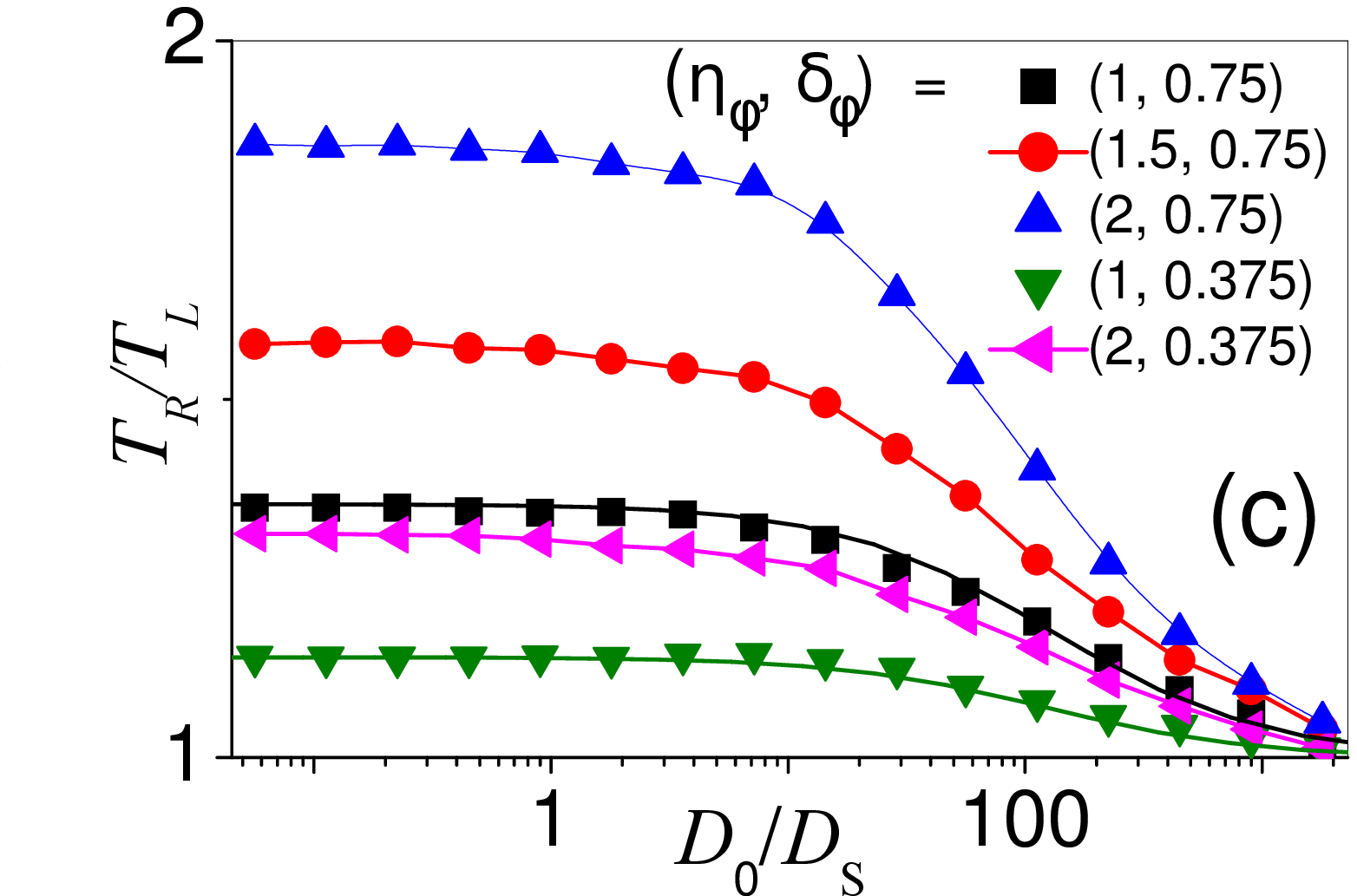}
\caption{(Color online) Channel diffusion for $\delta_v=0$ and
different $\delta_\phi$ and $\eta_\phi$ (in legends): (a) $N_R/N_L$
vs. $v_0$ for $D_0=0.03$; (b) $N_R/N_L$ vs $D_0$ for $v_0=1.5$; and
(c) $T_L/T_R$ vs $D_0$ for $v_0=1.5$. In (a)-(c) $D_\phi=0.1$, $L=100
l_\phi$, and $y_L=5$. The solid curves are the corresponding
analytical predictions based on Eqs. (\ref{Palpha})-(\ref{NRNL}) for
$\alpha=1$. \label{F4}}
\end{figure}

Regarding the transient statistics of a particle injected at the
center of the channel, $x=0$, a simple ``splitting probability''
calculation \cite{Risken} leads to
\begin{equation} \label{NRNL}
\frac{N_R}{N}=\frac{\int_{-L/2}^0[D_\alpha(x)P_0(x)]^{-1}dx}
{\int_{-L/2}^{L/2}[D_\alpha(x)P_0(x)]^{-1}dx}.
\end{equation}
with $N_R+N_L=N$. Analogously, for the MFPT's  in a channel with
absorbing endpoints, we have \cite{Risken}
\begin{equation} \label{TL}
T_L(\delta_\alpha)=\langle \tilde T(x) \rangle_{(0,L/2)}-\langle
\tilde T(x) \rangle_{(-L/2,L/2)}=T_R(-\delta_\alpha),
\end{equation}
where $\delta_{1/2}=\delta_v$ and $\delta_{1}=\delta_\phi$,
$$\tilde T(x)=\int_{-L/2}^x dz\psi_\alpha (z)/D_\alpha
(z)\int_z^{L/2}dy/\psi_\alpha(y),$$ and $\langle \dots\rangle_{(a,b)}
=\int_a^b (\dots)dx/\psi_\alpha(x)/\int_a^b dx/\psi_\alpha(x)$, with
$\psi_\alpha(x)=[D_\alpha(x)]^\alpha$. The second equality in Eq.
(\ref{TL}) follows immediately from $x \to -x$ symmetry
considerations. The ratios $N_R/N_L$ and $T_L/T_R$ have been computed
numerically. The results plotted for $\alpha=1/2$ (Fig. \ref{F3}) and
for $\alpha=1$ (Fig. \ref{F4}) confirm the consistency of our
phenomenological approach in both regimes.

Clearly, our approach hinges on the phenomenological Eq.~(\ref{LE2})
and the explicit expressions we used for $v_\alpha$ and
$D_\alpha(x)$. We now justify our choice for both.

{\it (i) $\delta_\phi=0$, $\delta_v> 0$}. In view of the third
equation (\ref{LE1}), we know that $\cos\phi(t)$ behaves like a
(non-Gaussian) colored noise with an asymptotic autocorrelation
function $\langle \cos \phi(t) \cos \phi(0)\rangle \simeq
(1/2)e^{-D_\phi |t|}$ for $t\gg \tau_\phi$ \cite{ourPRL}. The JP
diffusion coefficient at $x$ can thus be derived from Kubo's formula
\cite{EPJST,Marchesoni},
$$D=D_0 +\lim_{t\to \infty}\int_0^t v_0^2(x) \langle \cos \phi(t)
\cos \phi(0)\rangle dt=D_{1/2}(x),$$  as anticipated in Eq.
(\ref{LE2}).

The drift velocity, $v_\alpha(x)$, of a JP with an $x$-independent
self-propulsion time constant, $\tau_\phi$, amounts to the average of
$v_0(x)$ over its persistence length $l_\phi(x)=v_0(x)\tau_\phi$,
i.e.,
\begin{equation} \label{v12}
v_{\alpha}(x)=\frac{1}{2}\left [v_0\left(x+\frac{l_\phi}{2}\right
)-v_0\left(x-\frac{l_\phi}{2}\right )\right]\simeq
\frac{1}{2}v'_0(x)v_0(x)\tau_\phi,
\end{equation}
hence $v_{\alpha}=\alpha D'_\alpha(x)$ as in Eq.~(\ref{FPE}) with
$\alpha=1/2$.

{\it (ii) $\delta_\phi > 0$, $\delta_v=0$}.
%
%
Calculating $D(x)$ in this case is straightforward. The FPE
corresponding to the first and third Langevin equations (\ref{LE1}),
$$ \frac{\partial}{\partial t}\bar P=\left[ -v_0\cos \phi
\frac{\partial}{\partial x} +D_0 \frac{\partial^2}{\partial x^2}
+D_\phi(x) \frac{\partial^2}{\partial \phi^2}\right ]\bar P,$$ with
$\bar P=\bar P(x,\phi,t)$, admits a uniform p.d.f., as one can prove
by substitution; hence, the $P_0(x)$ of Eq. (\ref{Palpha}) with
$\alpha=1$. The diffusion coefficient will be calculated again
through Kubo's formula: since in the stationary regime $x$ and $t$
are
statistically independent, $D=D_1(x)$. 
Regarding the drift velocity, the condition $j_0=0$ in Eq.(\ref{LE2})
requires that $v_\alpha(x)= D_\alpha(x)(\ln[D_\alpha(x)P_0(x)])'$,
namely, for $\alpha=1$, $v_1(x)=D'_1(x)$, as expected.

Coming  back to the plots of Figs. \ref{F3} and \ref{F4}, we stress
that:

(i) The insets of Figs.~\ref{F3}(b) and (c) illustrate the dependence
of $N_{R,L}$ and $T_{R,L}$ on the channel length $L$: $T_{R,L}$ scale
like $L^2$, whereas $N_{R,L}$ grow insensitive to $L$. Of course,
both statements are valid as long as $L \gg l_\phi$;

(ii) Our expressions for $T_{L,R}(\delta_\alpha)$, adapted from Ref.
\cite{Risken}, correctly reproduce the limiting values
$T_{R,L}(\delta_\alpha)=L^2/8D_0$ for $D_0\ll D_s$ (gradient effects
are superseded by thermal noise), and $T_{R,L}(0)=L^2/8D_s$ for $D_0
=0$ and $\delta_\alpha \to 0$ (purely active Brownian motion);

(iii) On comparing the curves for $N_R/N_L$ in Figs. \ref{F3}(b) and
\ref{F4}(b) and those for $T_L/T_R$ in Figs. \ref{F3}(c) and
\ref{F4}(c), the different dependence of the two ratios on $D_\phi$
at low thermal noise becomes apparent. This can be easily explained
by inspecting the corresponding analytical expressions in the limit
$D_0 \to 0$. For $\delta_\phi=0$, $\delta_v
> 0$, i.e., $\alpha=1/2$, the two ratios are functions of
$\delta_v$ only and, therefore, independent of $D_\phi$. For
$\delta_v=0$, $\delta_\phi \neq 0$, i.e., $\alpha=1$, they grow
insensitive to $v_0$, but do depend on $\delta_\phi$ and, hence,
$D_\phi$. Accordingly, the limits $D_0 \to 0$ and $v_0 \to \infty$
coincide, as confirmed, for instance, by the numerical data in Figs.
\ref{F4}(a) and (b).

Finally, to fully answer our starting question (1), as the most
likely exit end in the general case $\delta_v \neq 0$ and
$\delta_\phi \neq 0$, we computed the rectification factor
\begin{equation} \label{eps}
\epsilon\equiv \langle v_\alpha (0)\rangle/v_0=\left
(\frac{N_R}{N_L}\frac{T_L}{T_R}-1
\right)\mathlarger{\mathlarger{\mathlarger{\mathlarger{/}}}}\left
(\frac{N_R}{N_L}\frac{T_L}{T_R}+1 \right),
\end{equation}
where the ratios $N_R/N_L$ and $T_L/T_R$ are combined together to
quantify both the sign and magnitude of the symmetry-breaking
mechanism responsible for the pseudo-chemotactic drifts. The most
effective right (left) rectification corresponds to $\epsilon = \pm
1$, whereas for $\epsilon =0$ the opposite pulls by $v_0(x)$ and
$D_\phi(x)$ cancel each other. Note that for $\eta_v=1$,
$\eta_\phi=2$ and $\delta_v=\delta_\phi$, the transient rectification
does not vanish [Fig. \ref{F2}(c)], even if under the same conditions
$T_L=T_R$ [Fig. \ref{F2}(b), inset].

\section{Concluding remarks} \label{conclusions}

The phenomenon of ``drift without current'' has been explained in
Refs.~\cite{Ostrowsky,Ripoll} using the phenomenological
Eq.~(\ref{LE2}), by noticing that the statistical ensemble governing
the average drift (i.e., the rectification factor $\epsilon$ in our
notation) is different from the one required to compute the average
current, $j_0=0$. The former consists of the representative points
exiting an infinitesimally narrow neighborhood with coordinate $x$,
with equal $x$-dependent jump length in either direction, whereas the
latter consists of all points crossing a channel cross-section with
coordinate $x$ at a given time, no matter what their jumping length.
The two ensembles differ as an effect of multiplicative noise [i.e.,
the $x$-dependence of $D_\alpha(x)$] and so do the currents thus
calculated.

Self-propelling artificial microswimmers reproduce that very same
situation as a combined effect of nonequilibrium and the higher
dimensionality of their dynamics, Eq. (\ref{LE1}). 
In contrast to bacterial chemotaxis \cite{Schnitzer}, for an
artificial microswimmer the self-propulsion parameters do not depend
on the orientation. Here a dependence on the swimmer's orientation
might come into play due to, say, inertial or memory (i.e., nonlocal)
effects, but surely not to some internal sensor-actuator pathways,
like in bacteria \cite{Berg}. The microswimmers considered here are
characterized by very low Reynolds numbers and small dimensions
compared to the $\rho(x)$ length scale; therefore, an orientation
dependence of the swimmer's self-propulsion mechanism is not an
option.

For artificial microswimmers under the most common experimental
conditions, $D_\phi$ is only weakly affected by the $\rho$ gradient,
while $v_0(x)$ is reported to grow linearly with $\rho$ and then
saturate at higher $\rho$ \cite{Gibbs,Ibele,Sen_propulsion,Lugli}.
The onset of ``anti-Fick'' cold-to-hot (pseudo-chemotactic) currents
can thus be easily demonstrated. For instance, a source steadily
releasing fuel into a JP suspension, causes a concentration gradient
around it; JP's with $x$-independent rotational dynamics are driven
away from the fuel source, whereas a tagged JP floating in such a
depletion zone actually drifts toward the source. This prediction is
in contrast with the experimental findings of Ref.~\cite{SenPRL},
where Au-Pt micro-rods are reported to progressively cluster around
an ${\rm H_2O_2}$ fuel source. If we assume that the self-propulsion
model of Eqs.~(\ref{LE1}) holds good for a free swimmer in the bulk
(as established under the most diverse experimental conditions
\cite{Gibbs,Bechinger,Takagi}), the only explanation for such a
discrepancy is that, upon migrating toward the fuel source, the JP's
come into contact with one another and eventually aggregate, as
suggested, e.g., in Ref. \cite{Marchetti}. Drifts without current
become observable at low swimmer concentration.

\section*{Acknowledgements} We thank RIKEN's RICC for computational resources.
Y. Li is supported by the NSF China under grant No. 11334007, and by
Tongji University under grant No. 2013KJ025. FN is partly supported
by the RIKEN iTHES Project, the MURI Center for Dynamic
Magneto-Optics, and a Grant-in-Aid for Scientific Research (S).

\end{document}